\documentstyle[aps,multicol,epsf,prl]{revtex}
\begin{document}
\title{Numerical Study of Spin and Chiral Order in a Two Dimensional $XY$ 
Spin Glass}
\author{J.M. Kosterlitz and N. Akino}
\address{Department of Physics, Brown University, Providence, RI 02912, USA}
\date{\today}
\maketitle
\draft
\begin{abstract}
The two dimensional $XY$ spin glass in is studied numerically by a finite 
size defect energy scaling method at $T=0$ in the vortex representation which 
allows us to compute the exact (in principle) spin and chiral domain wall 
energies. We confirm earlier predictions that there is no glass phase at 
any finite $T$. Our results strongly support the conjecture that 
both spin and chiral order have the same correlation length exponent 
$\nu_{s}=\nu_{c}\approx 2.70$. Preliminary results in $3d$ are also obtained.
\end{abstract}
\pacs{PACS numbers: 75.10.Nr, 05.70.Jk, 64.60.Cn}

\begin{multicols}{2}
The $XY$ spin glass has been the subject of considerable attention and 
controversy for some time and is still not understood. It has been known 
since the seminal work of Villain\cite{villain1} that vector spin glass 
models have chiral or reflection symmetry in addition to the continuous 
rotational symmetry. Consequently, the $XY$ spin glass may have two different 
glass orders, a spin glass order and a chiral glass order. It is widely 
accepted and has become part of the spin glass folklore
that, in two and three dimensions, chiral and spin variables decouple at long
distances and order independently\cite{KT1,KT2,K1,mg1} although there is a hint
that this may not hold in four dimensions\cite{jain}. Numerical estimates of 
the correlation length exponents $\nu$ in two dimensions, where both spin and 
chiral order set in at $T=0$ as $\xi_{s,c}\sim T^{-\nu_{s,c}}$, indicate 
that $\nu_{c} = 2.57\pm 0.003$ and $\nu_{s} = 1.29\pm 0.02$\cite{mg1} which 
agree with older, less accurate estimates\cite{KT1,KT2,rm}. The decoupling 
of chiral and spin degrees of freedom seems to be well established by these 
numerical results, but some analytic work on special models\cite{nhm,nh,tnh} 
implies that, at least for $XY$ spin glasses below their lower critical 
dimension $d_{l}>2$ when order sets in at $T=0$, {\it{both}} correlation 
lengths diverge with the {\it{same}} exponent $\nu_{s} = \nu_{c}$. To add 
to the confusion, there is rather convincing evidence that chiral order sets 
in for $0<T<T_{c}$, while spin glass order occurs only at $T=0$ in 
$3d$\cite{KT1,KT2,K1,jy,mcm,mcmbc}. These numerical investigations have led 
to another piece of accepted folklore, namely that the lower critical 
dimension $d_{l}\ge 4$ for spin glass order\cite{mcmbc,cb}. A very recent 
simulation\cite{mg1}
concluded that earlier simulations are misleading because the spin 
defect energy began to grow with system size $L$ at values of $L$ just 
beyond the limit accessible to earlier attempts and that $d_{l}$ is slightly 
less than three. However, chiral order is robust in $3d$. In $2d$, all 
simulations agree that chiral and spin glass order set in at $T=0$ but with 
different exponents $\nu_{c}\approx 2\nu_{s}\approx 2.6$.
\newline\indent
The theoretical situation is unclear since, to our knowledge, there is no 
unambiguous proof of any of the accepted folklore outlined 
above\cite{no1,sy,on1}, numerical simulations are contradictory\cite{mg1} 
and the analytic work on special models\cite{nhm,nh,tnh} is difficult 
to reconcile with the {\it{apparently}} unambiguous numerical simulations on 
the $2d$ $XY$ spin glass. In this letter, we attempt to clarify the 
contradictory conclusions from numerical and analytic studies outlined above
and to identify which should be retained and which need revision. Our 
essential conclusion is  that, by carefully defining spin and a chiral
domain wall energies, we find numerical agreement with the conjecture\cite{nh}
that $\theta_{s}=\theta_{c}$ in $2d$ where $\theta_{s,c} = -1/\nu_{s,c}$ are
the $T=0$ stiffness exponents. Although the conjecture is not rigorous, it
is the {\it{only}}, to our knowledge, analytic prediction existing and is the
only check we have on the validity or otherwise of the numerical method
used, at least until some rigorous testable predictions are made. If one 
accepts that a valid numerical simulation must agree with the conjecture, the
implications go far beyond minor points such as the numerical values of
stiffness exponents but implies that most of the $XY$ spin and gauge glass
folklore is incorrect. The lower critical dimension for {\it{both}} spin glass
and chiral order is $2<d_{l}<3$, the chiral glass scenario $\theta_{s}<0$ and
$\theta_{c}>0$ in $3d$ is not possible but both stiffness exponents are 
positive and the presently accepted numerical values in $2d$ and $3d$ are 
incorrect and need re-examination.
\newline\indent
A natural way of investigating order is to compute the domain wall or
defect energy $\Delta E(L)$ of a system of size $L$ for several 
realizations of disorder (samples) for different values of $L$ and fit to
the finite size scaling {\it{ansatz}}\cite{mcm,dwrg}
\begin{equation}
<\Delta E(L)>\sim L^{\theta_{s,c}}
\label{eq:fss}
\end{equation}
where $<\cdots>$ denotes an average over disorder, 
$\Delta E(L) =E_{D}(L)-E_{0}(L)$
the domain wall energy with $E_{0}(L)$ the ground state (GS) energy, 
$E_{D}(L)$ the energy of the system of size $L$ containing a spin or chiral 
domain wall and $\theta_{s,c}$ is the spin $(s)$ or chiral $(c)$ stiffness 
exponent. There
are two main difficulties in applying these ideas to a finite disordered 
system. The first is how to define $E_{0}$ and $E_{D}$ for a finite system 
with disorder since the GS configuration is unknown and the energy of a finite
system depends on the boundary conditions (BC) imposed which must be 
compatible with the GS configuration. A spin or chiral domain wall is induced 
by an appropriate change in these BC and $E_{D}$ is the minimum energy of the 
system subject to these new BC. The second difficulty is the computational
problem of finding $E_{0}$ and $E_{D}$ sufficiently accurately so the errors
in $\Delta E(L)$ can be controlled and kept small. The numerical data is
fitted to eq.(\ref{eq:fss}) in an attempt to verify the scaling ansatz and
to obtain numerical values of the fundamental stiffness exponents
$\theta_{s}$ and $\theta_{c}$. These constraints limit the accessible sizes 
$L$ to small values when the BC have large effects and it is essential to
treat the BC properly to define $E_{0}$ and $E_{D}$ consistently for a fit
of the numerical data to eq.(\ref{eq:fss}) to have any meaning.
\newline\indent
The Hamiltonian of a $\pm J$ $XY$ spin glass on a $L\times L$ square lattice is
\begin{equation}
\label{eq:h1}
H=\sum_{<ij>}V(\theta_{i}-\theta_{j}-A_{ij})
\end{equation}
where $V(\phi)$ is an even $2\pi$ periodic function of $\phi$ with a maximum 
at $\phi =\pi$, usually taken to be $V(\phi_{ij}) = -J_{ij}cos(\phi_{ij})$ 
with the coupling $J_{ij}=J>0$ for $ij$ nearest neighbor sites of 
a square lattice. The random bond variables $A_{ij} = 0,\pi$ with equal 
probability $1/2$ correspond to ferro and antiferromagnetic coupling 
between neighboring spins. We imagine the system of eq.(\ref{eq:h1}) on a 
torus which corresponds to imposing periodic BC on the phases 
$\theta_{i_{x},i_{y}} = \theta_{i_{x}+L,i_{y}} = \theta_{i_{x},i_{y}+L}$ 
with $i_{x,y}= 1,\cdots,L$ and
coupling spins on opposite faces by some interaction 
$\tilde{V}(\theta_{L,i_{y}},\theta_{1,i_{y}})$ and
$\tilde{V}(\theta_{i_{x},L},\theta_{i_{x},1})$ which may be regarded as 
defining the BC. In principle, the GS is obtained by minimizing the energy
with respect to the $L^{2}$ bulk variables $\theta_{i}$ {\it{and}} all
possible $\tilde{V}$. This program is too difficult for us so we will restrict
ourselves to $\tilde{V}$ which impose a spin defect or a chiral defect. To
impose a spin domain wall, we choose 
$\tilde{V} = V(\theta_{i}-\theta_{j}-A_{ij})$ where the phase differences 
$A_{ij}$ between corresponding sites $i,j$ on opposite faces may be varied to
find the minimum energy $E_{0}$. It is not necessary to vary every $A_{ij}$
as each elementary plaquette on the torus is equivalent and the plaquettes
between opposite faces are indistinguishable from the others and play no
special role. We therefore keep fixed the frustrations 
$f_{\bf r}=\sum_{\Box{\bf r}}A_{ij}/2\pi$ fixed where the sum is over the 
bonds in a clockwise direction of the elementary plaquette whose center
is at the site ${\bf r}$ of the dual lattice. We are free to choose
$\tilde{V}$ to impose a global phase twist $\Delta_{\mu}=0,\pi$ in
the direction $\mu$ round the torus. 
The lowest energy $E_{0}({\Delta_{\mu}})$ is $2\pi$ periodic 
in $\Delta_{\mu}$ with a minimum at some $\Delta^{0}_{\mu}$ which depends
on the particular sample. To introduce a spin domain wall perpendicular to $x$,
one simply changes the twists from their best twist (BT) values 
$\Delta^{0}_{\mu}\rightarrow(\Delta^{0}_{x}+\pi,\Delta^{0}_{y})$ and find the
minimum energy subject to this constraint, which yields the energy with a spin 
domain wall $E_{sD}(L)>E_{0}(L)$. The spin defect energy
$\Delta E^{BT}_{s}(L)\equiv E_{sD}(L)-E_{0}(L)$ is computed for
different samples and sizes $L$ and fitted to
\begin{equation}
<\Delta E^{BT}_{s}(L)>\sim L^{\theta^{BT}_{s}}
\label{eq:sdw}
\end{equation} 
to obtain the spin stiffness exponent $\theta^{BT}_{s}$. A chiral domain
wall is imposed by reflective BC\cite{K1,nh} which means that there is a
seam encircling the torus in (say) the $y$ direction across which the spins
interact as $\tilde{V} = V(\theta_{i}+\theta_{j}-A_{ij})$ which is 
equivalent to a
reflection of the spins about some arbitrary axis. In principle, one can
follow the same procedure as for the spin domain wall to obtain the
chiral defect energy $\Delta E^{BT}_{c}(L) = E_{cD}(L)-E_{0}(L)$ where 
$E_{cD}$ is the minimum energy with the modified interactions on a seam.
However, there is no reason to expect that $E_{cD}>E_{0}$ as the BC defining
$E_{0}$ may trap a chiral defect in some samples in which cases the modified
interactions cancel the chiral defect and $E_{cD}<E_{0}$, as in fact 
does occur. We therefore define $\Delta E_{c}^{BT} = \mid E_{cD}-E_{0}\mid $,
average this over disorder and fit to 
$<\Delta E_{c}^{BT}>\sim L^{\theta_{c}^{BT}}$ to obtain the chiral stiffness
exponent. This does not affect $\Delta E_{s}^{BT}$ as both $E_{sD}$ and 
$E_{0}$ contain the same chiral defects. The procedure described above 
using the phase representation of
eq.(\ref{eq:h1}) is similar to that of most previous 
studies\cite{KT1,KT2,K1,mg1,rm} except that these omit the minimization with
respect to the twists $\Delta_{\mu}$, apply naive periodic and antiperiodic BC 
and call the lowest energies $E_{p}$ and $E_{ap}$. Neither of these BC is
compatible with the GS and both must induce some excitation from $E_{0}$. 
Nevertheless, the spin defect energy is defined as 
$\Delta E_{s}^{RT}\equiv \mid E_{ap}-E_{p}\mid$ and the spin stiffness 
exponent defined by $<\Delta E_{s}^{RT}(L)>\sim L^{\theta_{s}^{RT}}$. We call
this a random twist (RT) measurement as both BC are equivalent to some
random choice of $\Delta_{\mu}$ relative to $\Delta^{0}_{\mu}$ for each
sample. There is no good reason to expect $\Delta E^{RT}_{s}(L)$ to scale as
$L^{\theta_{s}}$ but if it does, there is less reason to expect any
relation between $\theta_{s}^{RT}$ and $\theta_{s}^{BT}$ or $\theta_{c}$.
\newline\indent
The procedure in terms of the phase representation of the
$XY$ spin glass Hamiltonian of eq.(\ref{eq:h1}) is followed by previous
studies. The aim is to obtain $\Delta E(L)$ by independently minimizing
the Hamiltonian with respect to the $\theta_{i}$ to obtain $E_{D}$ 
and $E_{0}$. This requires finding essentially exact global minima for
each sample to control the errors in $<\Delta E(L)>$ to be purely
statistical and $O(N^{-1/2})$ where $N$ is the number of samples. If the
minimization algorithm fails to find the true global minima, the errors
in $<\Delta E(L)>$ will be uncontrolled and very large, making the data
point useless. Since the $\mid\theta_{i}\mid\leq\pi$ are continuous, one has to
perform a numerical search of a huge configuration space, most of which does
not even correspond to a {\it local} energy minimum. To reduce the volume
of the space, we transform to a Coulomb gas (CG) representation which 
eliminates spin wave excitations and parametrizes the problem in terms of
integer valued vortex or charge configurations, each of which is a local
energy minimum. This reduces the space to be searched to manageable size
although it introduces long ranged Coulomb interactions between
vortices.
The potential $V(\phi)$ in eq.(\ref{eq:h1}) is taken as a piecewise 
parabolic potential equivalent
to a Villain\cite{villain2} potential at $T=0$
\begin{eqnarray}
\label{eq:h2}
H &=& \frac{J}{2}\sum_{<ij>}(\theta_{i}-\theta_{j}-A_{ij}-2\pi n_{ij})^{2}
\cr
   &\equiv &\frac{J}{2}\sum_{<ij>}(\phi_{ij}-A_{ij})^{2}
\end{eqnarray}
where $n_{ij}=-n_{ji}$ is any integer on the bond $ij$. By a 
duality transformation\cite{nh,jkkn,vb} the Coulomb gas Hamiltonian  
with periodic BC in the phases becomes
\begin{eqnarray}
\label{eq:hcg}
H = & 2&\pi^{2}J\sum_{{\bf r},{\bf r'}}(q_{\bf r}-f_{\bf r})
G({\bf r}-{\bf r'})(q_{\bf r'}-f_{\bf r'})
\cr
   & +&J(\sigma_{x}^{2}+\sigma_{y}^{2})/2L^{2}
\end{eqnarray}
where
\begin{eqnarray}
\label{eq:defs}
\sigma_{x} &=& -2\pi[L(q_{x1}-f_{x1})+\sum_{\bf r}(q_{\bf r}-f_{\bf r})y]  \cr
\sigma_{y} &=& -2\pi[L(q_{y1}-f_{y1})-\sum_{\bf r}(q_{\bf r}-f_{\bf r})x]   \cr
G({\bf r}) &=& \frac{1}{L^{2}}\sum_{{\bf k}\neq 0}\frac{e^{i{\bf k\cdot r}}-1}
{4-2cosk_{x}-2cosk_{y}}
\end{eqnarray}
Here, ${\bf r}=(x,y)$ denotes the sites of the dual lattice and $G({\bf r})$ 
is the lattice Green's function. In eq.(\ref{eq:defs}), 
$k_{\alpha} =2\pi n_{\alpha}/L$ with $n_{\alpha} = (0,1,\cdots, L-1)$.
The topological charge, $q_{\bf r}$, is the circulation of the phase 
about the plaquette at ${\bf r}$ and can be any
integer subject to the neutrality condition $\sum_{\bf r}q_{\bf r}=0$. The 
frustration at ${\bf r}$, $f_{\bf r} = \sum_{\Box{\bf r}}A_{ij}/2\pi$, is the 
circulation of $A_{ij}$ round the plaquette. 
$f_{x1}=\sum_{\Box x}A_{ij}/2\pi$ is the circulation 
round the whole torus on the $x$ bonds of plaquettes at $y=1$
and $q_{x1}$ is the circulation of the phase. $f_{y1}$ and 
$q_{y1}$ are defined similarly. Periodic BC in the phases $\theta_{i}$ 
restrict $q_{x1},q_{y1}$ to be integers.      
A chiral domain wall is introduced by reflective BC when the
Hamiltonian becomes\cite{nh} 
\begin{eqnarray}
\label{eq:href}
H_{R}&=&2\pi^{2}J\sum_{{\bf r},{\bf r'}}(q_{\bf r}-f_{\bf r})
(q_{\bf r'}-f_{\bf r'})G_{R}({\bf r}-{\bf r'})
\cr  
G_{R}({\bf r})&=&\frac{1}{L^{2}}\sum_{\bf \kappa}
\frac{e^{i{\bf \kappa\cdot r}}}{4-2cos\kappa_{x}-2cos\kappa_{y}}
\end{eqnarray}
where $\kappa_{x}=\pi(2n_{x}+ 1)/L$ and $\kappa_{y}=k_{y}$ so that 
$G_{R}({\bf r})= G_{R}({\bf r}+L{\bf\hat y}) = -G_{R}({\bf r}+L{\bf\hat x})$ 
and the charges $q_{\bf r}$ obey 
a modified neutrality condition 
$(\sum_{\bf r}q_{\bf r}+2f_{1y}){\text{mod}} 2=0$\cite{nh}. A 
more convenient form of the Hamiltonian for simulation purposes is by 
doubling the lattice in
the $x$ direction to a $2L\times L$ lattice in which the extra half is a 
charge conjugated image of the original so that
\begin{equation}
\label{eq:hdouble}
H_{R}=\pi^{2}J\sum_{{\bf r},{\bf r'}}(q_{\bf r}-f_{\bf r}){\tilde{G}}
({\bf r}-{\bf r'})(q_{\bf r'}-f_{\bf r'})
\end{equation}
where ${\tilde{G}({\bf r})}$ is the Green's function for a $2L\times L$ 
lattice with {\it{periodic}} BC and $q_{{\bf r}+L{\bf\hat x}}=-q_{\bf r}$,  
$f_{{\bf r}+L{\bf\hat x}} = -f_{\bf r}$\cite{nh}.
\newline\indent
To estimate the spin stiffness exponent $\theta_{s}$, simulations were 
performed on a $L\times L$ lattice with eq.(\ref{eq:hcg}) in two different 
ways. The first is by a RT measurement by imposing standard periodic
and antiperiodic BC corresponding to $\Delta_{x}=0$ and $\Delta_{x}=\pi$, then 
fitting to $<\Delta E^{RT}>\sim L^{\theta}$. This is just the procedure 
followed by all previous studies and, not
surprisingly, gives essentially the same result 
$\theta_{s}^{RT}=-0.76\pm 0.015$\cite{KT1,KT2,K1,mg1,rm} with system sizes
$L=4,5,6,7,8,10$ and averaging over $2560$ samples for $L\leq 8$ and $1152$ 
for $L=10$ (see Fig.(1)). This way 
of measuring a spin domain wall energy does not exploit all the freedom 
implied by eq.(\ref{eq:hcg}). One can find the global energy minimum by 
optimizing the BC by allowing the combinations $(q_{x1}-f_{x1})$ and
$(q_{y1}-f_{y1})$ to vary independently over any integer or half integer. 
This corresponds to allowing the circulations of the phase difference and 
of $A_{ij}$ round the two independent loops encircling
the torus to vary. The absolute minimum energy $E_{0}$ is the GS energy 
(of a particular sample) and a spin domain wall is induced by 
$f^{0}_{x1}\rightarrow f^{0}_{x1}+1/2$. The energy minimum
$E_{sD}$ with these BC includes the energy due to the spin domain wall. 
Fitting the difference, $\Delta E_{s}^{BT}(L)\ge 0$, to eq.(\ref{eq:sdw}) 
yields $\theta_{s}^{BT}=  -0.37\pm 0.015$, averaging
over the same number of samples as in the RT measurement. We call this
a best twist (BT) measurement. This is equivalent to making a gauge 
transformation to all bonds in the direction $\mu =(x,y)$ by 
$A_{ij}\rightarrow A_{ij}+\Delta_{\mu}/L$. The energy $E$ is $2\pi$
periodic in $\Delta_{\mu}$, 
$E(\Delta_{\mu}) =E(\Delta_{\mu}+2\pi)$ and has a minimum at
some $\Delta_{\mu}^{0}$ which depends on the particular realization of 
disorder. The RT measurement keeps $f_{x1}$ fixed or $\Delta_{\mu}=0$, calling 
the lowest energy $E_{p}$, then changing $f_{x1}\rightarrow f_{x1}+1/2$ and 
calling the resulting lowest energy $E_{ap}$ and assuming the energy 
difference scales as $L^{\theta_{s}^{RT}}$. This procedure is equivalent to 
choosing an arbitrary gauge $A_{\mu}({\bf r})$ to compute $E_{p}$ and then 
$E_{ap}$ is computed in the gauge $A_{\mu}+\pi\delta_{\mu,x}/L$. The original 
problem of eq.(\ref{eq:h1}) is invariant under discrete gauge transformations 
${\text{modulo}}\;2\pi$ so the RT measurement is performed in a {\it{random}} 
gauge while the BT measurement is done in the gauge which
minimizes the energy and depends on the realization of disorder. We use 
simulated annealing\cite{anneal1,anneal2} to estimate the energy minima, 
which is much more efficient than simple quenching to $T=0$.
\newline\indent
The chiral domain  wall energy is also measured in two ways.
Defining $<\Delta E^{RT}_{c}>\equiv <|E_{m}-<E_{m}>|>$ \cite{KT2} 
where $E_{m}= {\text{min}}(E_{p},E_{ap}) -E_{R}$ 
with $E_{R}$ the GS energy with reflective BC gives the RT 
measurement for $\Delta E_{c}^{RT}$ 
and we obtain $\theta^{RT}_{c} =-0.37\pm 0.015$. The other way is the BT 
measurement which is analogous
to that for $\theta^{BT}_{s}$ when the absolute minimum energy is when 
the boundary terms in eq.(\ref{eq:hcg}) vanish. Since the the lowest energy 
of eq.(\ref{eq:hdouble}) may contain a chiral
but not a spin domain wall, the BT condition will hold and any boundary 
terms must vanish. Even if, in general, there were boundary contributions to 
eq.(\ref{eq:hdouble}), they would vanish in the BT condition. Thus, a BT 
measurement of $\Delta E^{BT}_{c}$ is obtained from
$\mid E^{BT}_{R} -E^{BT}_{0}\mid$ where $E^{BT}_{R}$ is the minimum of 
eq.(\ref{eq:hdouble}) and $E^{BT}_{0}$ is the minimum of eq.(\ref{eq:hcg}). 
Fitting to $<\Delta E^{BT}_{c}(L)>\sim L^{\theta_{c}^{BT}}$
yields $\theta^{BT}_{c} = -0.37\pm 0.010$. This implies that 
$\theta_{c}^{BT} =\theta_{s}^{BT} \approx -0.37$ to within numerical 
accuracy, agreeing with the conjecture of Ney-Nifle and Hilhorst\cite{nh}.
Note that the value of $\theta^{RT}_{s} \approx -0.76$ does not satisfy the 
conjecture. The only difference between the
RT and BT measurements is in $E_{0}$ from eq.(\ref{eq:hcg}) where 
$E_{0}^{RT}$ is obtained with fixed random BC and $E_{0}^{BT}$ by also 
minimizing with respect to the BC. $E_{cD}^{BT}$ and $E_{cD}^{RT}$
are both obtained from eq.(\ref{eq:hdouble}) and are identical because this 
is automatically a BT measurement for the special case of the spin glass as 
the boundary contributions to the energy vanish. Note that 
both measurements give identical values for the 
chiral exponent $\theta_{c}$ to within numerical uncertainty while the spin 
stiffness exponents $\theta_{s}^{BT}$ and $\theta_{s}^{RT}$ differ by a 
factor of two. All $2d$ results are in Fig.(1).
\newline\indent
Since the numerical estimates of $\theta_{s}^{BT}$ and $\theta_{c}^{BT}$
agree with the crucial test in $2d$\cite{nh}, we can regard this as
supporting our contention that we have a good definition of the defect
energies and our numerical method is fairly accurate. We have done
simulations on the $3dXY$ spin glass to estimate the spin stiffness
exponent $\theta_{s}$ and find $\theta_{s}^{BT} =+0.10\pm 0.04$ with
$L=2,3,4,5$ (Fig.(2)). This is larger and more accurate than the estimate
of ref.\cite{mg1}. The large error is due to fitting over only $3$ data
points. The negative slope of $\Delta E_{s}^{RT}(L)$ for $L=2,3,4$ is
expected to become positive at larger $L$\cite{mg1}. At present, we have
been unable to derive the $3d$ analogues of 
eqs.(\ref{eq:href},\ref{eq:hdouble}), so we have no estimate of
$\theta_{c}$\cite{mg1} in $3d$. 
\newline\indent
Computations were performed at the Theoretical Physics Computing Facility at
Brown University. JMK thanks A. Vallat for many discussions on $XY$ spin
glasses and on the importance of the CG representation when seeking the
ground state.

\begin{figure}
\label{fig1}
\center
\begin{minipage}{8.0cm}
\epsfxsize= 8.0cm \epsfbox{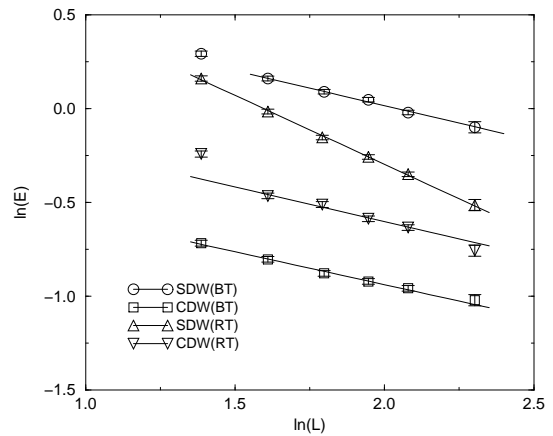}
\caption{Top to bottom: $L$ dependence of $\Delta E_{s}^{BT}$, 
$\Delta E_{s}^{RT}$, $\Delta E_{c}^{RT}$, $\Delta E_{c}^{BT}$ for 
$L=4,5,6,7,8,10$. Solid lines are power law fits.}
\end{minipage}
\end{figure}

\begin{figure}
\label{fig2}
\center
\begin{minipage}{8.0cm}
\epsfxsize=8.0cm \epsfbox{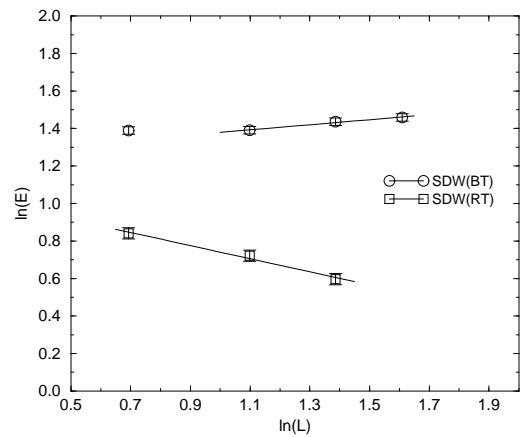}
\caption{$\Delta E_{s}^{BT}$ and $\Delta E_{s}^{RT}$ in $3d$.
Solid lines are power law fits.}
\end{minipage}
\end{figure}
 
\end{multicols}
\end{document}